# The Mechanism of Tidal Triggering of Earthquakes at Mid-Ocean Ridges

Christopher H. Scholz[1*], Yen Joe Tan[1], and Fabien Albino[2]

**Evidence for the triggering of earthquakes by tides has been largely lacking for the continents but detectable in the oceans where the tides are larger. By far the strongest tidal triggering signals are in volcanic areas of mid-ocean ridges. These areas offer the most promise for the study of this process, but even the most basic mechanism of tidal triggering at the ridges has been elusive. The triggering occurs at low tides, but as the earthquakes are of the normal faulting type, low tides should inhibit rather than encourage faulting. Here, treating the most well documented case, Axial Volcano on the Juan de Fuca ridge, we show that the axial magma chamber inflates or deflates in response to tidal stresses and produces Coulomb stresses on normal faults opposite in sign to those produced by the tidal stresses. If the bulk modulus of the magma chamber is below a critical value, the magma chamber Coulomb stresses will exceed the tidal ones and the phase of tidal triggering will be inverted. The stress dependence of seismicity rate agrees with triggering theory with unprecedented faithfulness, showing that there is no triggering threshold.**

Attempts to find evidence for tidal triggering of continental earthquakes have been largely negative or marginal[1-3]. However, in the oceans, where ocean loading can increase tidal stresses by an order of magnitude above the solid earth tides, there have been some successes. Cochran et al.[4] used a global catalog of oceanic earthquakes to show that shallow thrust earthquakes may be found to correlate with maximum tidally generated Coulomb stresses when the tides are large enough. Much stronger tidal triggering has been observed with ocean bottom seismometer networks in magmatic areas at mid-ocean ridges[5-9]. These are the most promising places to test theories of earthquake triggering. In these cases, however, even the most basic mechanism of the triggering is not understood. The most well studied of these is at Axial Volcano on the Juan de Fuca ridge. We shall study this case, and at the end, see if the results obtained there can also be applied to the others.

[1] Lamont-Doherty Earth Observatory, Columbia University, Palisades, NY, 10964, USA
[2] School of Earth Sciences, University of Bristol, Bristol, B58 1RJ, UK

Axial Volcano, which is at the intersection of a mid-ocean ridge with a hotspot, erupts on a decadal time scale. Each eruption is followed by caldera collapse accompanied by thrusting on outwardly-dipping ring faults, followed by a re-inflation period, at the latter stages of which the ring faults become reactivated in normal faulting[10-12]. The best observations of tidal triggering were for the normal faulting earthquakes in the months prior to the 2015 eruption[6].

At Axial Volcano the ocean tides are very large (3 m) so that ocean loading dominates the solid earth tides and the vertical tidal stress dominates and is in phase with the ocean tides (Supplementary Fig. S1), so we need only to consider the vertical component in our analysis. Tension is taken as positive for tidal stresses, so the maximum tidal stress corresponds to the minimum water depth. To avoid ambiguity, in this paper we will refer to high and low tides in the conventional way as high and low water, recalling that low water produces tension and high water compression.

Fig.1 shows a cross-section view of the seismicity prior to the 2015 eruption, which illuminates the ring faults. Fig. 2 shows a histogram of the seismicity plotted as a function of tidal period, in which 0° is the maximum low tide. The correlation is obvious and requires no statistical treatment. It was first proposed that this was a case of fault unclamping[6,8,9] but when it was established that these earthquakes were dominated by normal faulting[11] this viewpoint became untenable. Both the seismicity trends in Fig. 1 and the focal mechanisms[11] indicate a mean fault dip of 67°. A reduction of vertical stress brought about by low tide will produce a Coulomb stress change on such steeply dipping normal faults that inhibits their slip. It is, rather, the high tides that will produce a Coulomb stress on the faults that encourages slip. This seeming paradox is resolved by including the effect of the axial magma chamber on the distribution of stress.

**The response of the magma chamber**

The red curve in Fig. 1 delineates the roof of the axial magma chamber obtained from seismic imaging[13]. Inflation of the magma chamber drives the normal faulting on the ring faults. This is demonstrated in Fig. 3A, where we show the Coulomb failure stress change, $\Delta CFS = \Delta\tau - \mu\Delta\sigma$, on 67° dipping faults that results from a magma chamber overpressure of 1 MPa ($\Delta\tau$ is the change in shear stress resolved on the fault in the slip direction, $\Delta\sigma$ is the change in normal stress on the fault plane, and $\mu$ is the friction coefficient). Positive $\Delta CFS$ values encourage fault slip, negative ones inhibit it. The primary features in Fig. 3A are the zones of positive $\Delta CFS$ that correspond to the seismicity shown in Fig.1. See 'Methods' for details about the model.

     Because the magma chamber is a soft inclusion, its presence will profoundly affect the stress field in its vicinity resulting from any external load. We simulate the response to tides by calculating the distribution of $\Delta CFS$ on 67° dipping faults resulting from a reduction in vertical stress corresponding to a 1 m drop in the ocean tide. This is shown in Fig. 3B. The pattern is very similar to that of Fig. 3A, demonstrating how a

low tide can stimulate activity on these faults.  This pattern arises because the reduction of vertical stress causes the magma chamber, owing to its higher compressibility, to inflate relative to the surrounding rock, which produces a stress field congruent with that of Fig. 3A.  This is superimposed on a uniform $\Delta CFS$ from the tidal stress, which is negative in the case of a low tide.  Likewise, high tides cause the magma chamber to deflate, which also produces Coulomb stresses opposite in sign to the tidal ones.  Which component is larger determines whether earthquakes are stimulated by the low tide or the high tide.

The relative expansion of the magma chamber depends inversely with $K_m/K_r$, the bulk modulus of the magma relative to that of the surrounding rock, so this is the critical parameter that determines the behavior of the system.  In the calculation of Fig. 3A we used $\mu=0.8$, but because tidal loading is under undrained conditions, for calculations such as shown in Fig. 3B we use an effective friction $\mu'=(1-B)\mu$, where we adopted 0.5 for the value of Skempton's coefficient $B$. In Fig. 3B we used $\mu'=0.4$, $K_m=1$GPa and $K_r=55$GPa.  The systematics of the system are shown in Fig. 4 for several values of $\mu'$ and a constant $K_r=55$ GPa.  There the metric on the vertical axis, $\chi$, is the $\Delta CFS$ on a 67° dipping fault averaged from the corner of the magma chamber to the surface, normalized by the vertical tidal stress.  This is plotted against the bulk modulus of the magma.  Positive $\chi$ values indicate that earthquakes will be favored by low tides, negative values by high tides.  All conditions within the red region therefore favor earthquakes triggered on the low tide and inhibited on the high tide, and within the blue region, vice versa. The point indicated by the cross in Fig. 4 is the case illustrated in Fig. 3B.  The bulk modulus of gas-free magma is 12 GPa[14], but at the pressure of the magma chamber (~40MPa) this value can be reduced by one to two orders of magnitude by the presence of volatiles[15]. Thus, at this pressure, a magma of $K_m$ = 1GPa would contain 2650 ppm $CO_2$ by weight[12]. This is greater than the highest values typically seen for $CO_2$ content of MOR magma[16], but this difference could easily be accounted for by the inclusion of exsolved $H_2O$. So, we consider 1GPa to be a realistic value for $K_m$. We will take this choice of parameters as representative. They indicate $\chi$ = 0.32, a figure that will enter into the modeling calculations of the triggered seismicity in the next section.

**Modeling the earthquake triggering**

There are two models that relate change in seismicity rate to a rapid change in driving stress. These are based on earthquake nucleation models[17], one derived from the rate and state friction law[18] and the other from subcritical crack growth due to stress corrosion[19]. The rate-state friction version is

$$\frac{R}{r} = \exp\left[\frac{\Delta CFS}{A\sigma}\right] \qquad (1)$$

and the stress corrosion version is

$$\frac{R}{r} = \left[1 + \left(\frac{\Delta CFS}{\Delta \tau}\right)^n\right] \qquad (2)$$

where $R$ is the instantaneous seismicity rate, $r$ is the background rate, here taken as the rate when the tidal stress is zero, and $\Delta CFS = \chi \sigma_v$, the latter being the vertical tidal stress. The control parameters for the rate state friction version are the normal stress $\sigma$ and the 'viscous' friction term $A$. For the stress corrosion version, they are the stress corrosion index $n$ and the earthquake stress drop $\Delta \tau$.

The fit of these equations to the data is shown in Fig. 5, where the solid blue curve and the dashed red curves are eqn. (1) and (2), respectively. These two formulations cannot be distinguished and fit the data equally well. There is no detectable phase shift between the seismicity and the tides (Fig. 2), nor is there any hysteresis observed – data for rising and falling stresses fit the triggering curves equally well (Supplemental Fig. S2). We conclude that poroelastic relaxation is negligible in the response to the semi-diurnal tides.

The degree of conformity of data to the models shown in Fig. 5 is unprecedented. The various implications of this will be deferred to the discussion section.

**Applications to other areas**

Wilcock[5] searched for tidal triggering on the mid-ocean ridge systems of the NE Pacific, using mainly land-based networks. He found a 15% increase in seismicity within 15° of the lowest tides. The focal mechanisms of the earthquakes, however, were undetermined. With an OBS deployment on the Endeavour segment of the Juan de Fuca ridge, some 2° NE of Axial Volcano, the correlation of seismicity with low tides became much better defined[6]. Most of the triggered seismicity there was near the ridge axis, where the focal mechanisms indicate normal faulting[20]. This situation is therefore quite similar to Axial Volcano and the same triggering mechanism seems applicable.

At the hydrothermal field at 9°50′N on the East Pacific rise, an OBS deployment also showed evidence for tidal triggering[7]. There the ocean tides are much smaller than at Axial Volcano and a significant contribution to tidal stresses is made by the solid earth tides. The seismicity maximum correlates with the maximum extensional tidal stress, which can reach 1.3 kPa. The dependence of the seismicity on stress is similar to that observed at Axial Seamount (compare Fig.3c in ref. 7 to our Fig. 5). Evidence for the mechanism of the earthquakes is equivocal: scant focal mechanism data has indicated strike-slip, normal faulting and reverse faulting[21,22], and others have proposed that the seismicity it due to hydrothermally induced extension cracking[23]. There is also a variation in the tidal phase angle of earthquakes along the strike of the ridge axis. This indicates the earthquake triggering is also modulated by pore pressure changes brought about by hydrothermal circulation[24]. With this degree of ambiguity, we cannot assess how our deformation mechanism may be related to the tidal triggering in this location.

The unloading model used here was initially tested at Katla volcano (Iceland), where earthquakes show an annual cycle with the maximum seismicity rate occurring in the late summer[25] when the snow cover of the glacier above the volcano is minimum (annual fluctuation – 6m). The model[26] showed that this was also the period of maximum Coulomb stresses in the area above the magma chamber. However, in this case, it was not possible to correlate high Coulomb stress changes with the seismic events, because the focal mechanisms and the geometry of the faults were not known (P. Einarsson, pers. comm., 2018). At Axial Seamount, we have better constraints on the faulting system and our results show that low ocean tides can produce either reverse faulting or normal faulting above the magma chamber, depending on the magma compressibility (whether the system is within the blue or red areas of Fig. 4).

**Discussion**

Our observations of seismicity rate change as a function of stress, shown in Fig. 5, are unprecedented both in their breadth and faithfulness to the triggering models. The goodness of this fit is independent of the parameters in our magma chamber deformation model. The $\chi$ parameter, which incorporates those, affects only the scale of the stress axis, which determines the values of the control parameters. In the rate/state friction version, the representative value $\chi=0.32$ yields $A\sigma=0.0043$ MPa, about an order of magnitude smaller than found in earlier studies[4,27,28]. In the earlier studies the earthquakes were deeper (8-20 km), so the difference could be from that factor alone. In those papers, to accommodate lab values for *A* of 0.003-0.007, near-lithostatic pore pressures were assumed to get low enough values of $\sigma$ to match the observed $A\sigma$. At Axial Volcano the normal stress at the average earthquake depth of 1.2 km, assuming $\mu$=0.8, a hydrostatic pore pressure gradient, and dip 67°, is 7.2 MPa. It is not credible that overpressures can be maintained in the top 1 km of very young oceanic crust where there is no sediment cover and there is vigorous hydrothermal circulation throughout the caldera[29,30]. Using 7.2 MPa for $\sigma$ we conclude that *A*=0.0006, much smaller than lab values. Considering the entire spread of the solution space for $\mu'$= 0.4, $\chi$ ranges from 0.42 to zero, so with $\sigma$ = 7.2 MPa, the equivalent range of *A* is 0.0007≥*A*>0. If hydrostatic pore pressure was assumed in the other studies, estimates for *A* in that range would be obtained. The higher sensitivity to triggering at Axial Volcano is due to the shallow depths of earthquakes there. Considering the Vidale et al.[2] study, which failed to detect a tidal correlation for Southern California earthquakes, if we use 8 km depth with a hydrothermal pore pressure gradient, their typical tidal stress level of 1 kPa, and our *A* value of 0.0006, we calculate[31] that they would need 90k events to detect a correlation, whereas their catalog contained only 13k. Thus, the other studies are consistent with our finding that the *A* parameter at geologic rates must be considerably smaller than lab values. Laboratory studies often show that the friction rate parameters *A* and *B* depend on sliding rate[32,33]: the few experiments at plate tectonic slip rates[34] indicate that the friction parameters at those rates may differ significantly from those measured at the much higher rates usually employed in laboratory experiments.

Beeler and Lockner[31] noted that there are two triggering regimes: a threshold regime, in which the earthquake nucleation time $t_n$ is shorter than the tidal period and a nucleation regime, in which it is longer. In the former, maximum seismicity rate would correlate with the maximum stressing rate, in the latter, with maximum stress amplitude. Our data clearly confirm the latter (Fig.2), and the latter is also implicit in the fit in Fig. 5. The uplift rate prior to the 2015 eruption was 61 cm/yr[12]. From our inflation model (e.g. Fig. 3A) we find that the corresponding fault stressing rate $\dot{\tau}$ is 5 MPa/yr. Using[31] $t_n = \frac{2\pi A\sigma}{\dot{\tau}}$, we get $t_n$= 48 hrs. confirming that the system is indeed in the nucleation regime.

For the stress corrosion version of the triggering equation, if we take the stress corrosion index to be the laboratory values for basalt, 22<$n$<44[35], then the best fitting stress drop would be 0.09<$\Delta\tau$<0.17 MPa. This is a bit lower than the 0.18<$\Delta\tau$<2.8 range[35] for earthquakes at 1 km depth in Southern California, although these estimates are from mainly strike-slip earthquakes, which have systematically higher stress drops than normal faults[36]. If we take the rule that stress drop is about 3% of the shear strength[37], then for strength $\tau=\mu\sigma$ = 5.7 MPa we get $\Delta\tau$=0.17 MPa, within the range of our fit value. Thus, for this version of the triggering law, we do not have any serious conflict with independent estimates.

Thresholds for static or dynamic triggering have been much discussed[38-40]. Van der Elst and Brodsky[41] showed that dynamic triggering could be detected at very small strains, and suggested that the lower limit may simply be a matter of detectability. Our results (Fig. 5) show that seismicity rate falls smoothly as the tidal stress falls to zero, indicating that there is no threshold for triggering. Seismicity rate continues to fall when the tidal Coulomb stress becomes negative, indicating that what is often called 'stress shadowing' is a continuous quantifiable function of stress reduction.

It has often been remarked that hydrothermal areas seem particularly susceptible to dynamic triggering from distant earthquakes[12-44]. Attempts to explain this have invoked various effects of dynamic stresses on the permeability and/or pressure of the pore fluid[45-47]. The excellent agreement of our data with the 'dry' triggering models indicates that additional mechanisms are not required to explain the tidal triggering at Axial Volcano. In the case of tidal triggering some of those proposed mechanisms, such as unclogging of fluid pathways, are less likely because the tides are continually jostling the faults so that clogs, such as from mineralization, as suggested in the Yellowstone case[43], will not have time to form.

**Acknowledgements**

Mike Burton is thanked for instructing us on the properties of magmas at mid-ocean ridges. The earthquake catalog is the same as in Wilcock (2016) and is archived in the Interdisciplinary Earth Data Alliance Marine Geoscience Data System (DOI: 10.1594/IEDA/322843). That data was collected with funding from NSF under grant OCE-1536320.

**Author contributions**

CHS conceived the main ideas, led the project, and wrote the manuscript. YJT performed the analysis of the tides and earthquake data, fit the triggering equations, contributed to their interpretation, and prepared Figs. 1,2, and 5. FA developed the deformation model, carried out all deformation modeling, and prepared Figs. 3 and 4.

**Competing Interests**

The authors declare no competing interests.

1       Hartzell, S. & Heaton, T. The fortnightly tide and the tidal triggering of earthquakes. *Bulletin of the Seismological Society of America* **79**, 1282-1286 (1989).
2       Vidale, J., Agnew, D., Johnston, M. & Oppenheimer, D. Absence of earthquake correlation with earth tides: an indication of high preseismic fault stress rate. *J. Geophys. Res.* **103**, 24,567-524,572 (1998).
3       Wang, W. & Shearer, P. M. No clear evidence for localized tidal periodicities in earthquakes in the central Japan region. *Journal of Geophysical Research-Solid Earth* **120**, 6317-6328, doi:10.1002/2015jb011937 (2015).
4       Cochran, E. S., Vidale, J. E. & Tanaka, S. Earth tides can trigger shallow thrust fault earthquakes. *Science* **306**, 1164-1166, doi:10.1126/science.1103961 (2004).
5       Wilcock, W. S. D. Tidal triggering of earthquakes in the Northeast Pacific Ocean. *Geophysical Journal International* **179**, 1055-1070, doi:10.1111/j.1365-246X.2009.04319.x (2009).
6       Wilcock, W. S. D. *et al.* Seismic constraints on caldera dynamics from the 2015 Axial Seamount eruption. *Science* **354**, 1395-1399, doi:10.1126/science.aah5563 (2016).
7       Stroup, D. F., Bohnenstiehl, D. R., Tolstoy, M., Waldhauser, F. & Weekly, R. T. Pulse of the seafloor: Tidal triggering of microearthquakes at 9 degrees 50'N East Pacific Rise. *Geophysical Research Letters* **34**, doi:10.1029/2007gl030088 (2007).
8       Wilcock, W. S. D. Tidal triggering of micro earthquakes on the Juan de Fuca Ridge. *Geophysical Research Letters* **28**, 3999-4002, doi:10.1029/2001gl013370 (2001).
9       Tolstoy, M., Vernon, F. L., Orcutt, J. A. & Wyatt, F. K. Breathing of the seafloor: Tidal correlations of seismicity at Axial volcano. *Geology* **30**, 503-506, doi:10.1130/0091-7613(2002)030<0503:botstc>2.0.co;2 (2002).
10      Chadwick, W. W., Jr., Nooner, S. L., Butterfield, D. A. & Lilley, M. D. Seafloor deformation and forecasts of the April 2011 eruption at Axial Seamount. *Nature Geoscience* **5**, 474-477, doi:10.1038/ngeo1464 (2012).
11      Levy, S. *et al.* Mechanics of fault reactivation before, during, and after the 2015 eruption of Axial Seamount. *Geology* **46**, 447-450, doi:10.1130/g39978.1 (2018).
12      Nooner, S. L. & Chadwick, W. W. Inflation-predictable behavior and co-eruption deformation at Axial Seamount. *Science* **354**, 1399-1403, doi:10.1126/science.aah4666 (2016).
13      Arnulf, A. F. *et al.* Anatomy of an active submarine volcano. *Geology* **42**, 655-658, doi:10.1130/g35629.1 (2014).
14      Johnson, D. J., Sigmundsson, F. & Delaney, P. T. Comment on "Volume of magma accumulation or withdrawal estimated from surface uplift or subsidence, with application to the 1960 collapse of Kilauea volcano" by P. T. Delaney and D. F. McTigue. *Bulletin of Volcanology* **61**, 491-493, doi:10.1007/s004450050006 (2000).
15      Huppert, H. E. & Woods, A. W. The role of volatiles in magma chamber dynamics. *Nature* **420**, 493-495, doi:10.1038/nature01211 (2002).
16      Le Voyer, M., Kelley, K. A., Cottrell, E. & Hauri, E. H. Heterogeneity in mantle carbon content from $CO_2$-undersaturated basalts. *Nature Communications* **8**, doi:10.1038/ncomms14062 (2017).

17  Kanamori, H. & Brodsky, E. E. The physics of earthquakes. *Reports on Progress in Physics* **67**, 1429-1496, doi:10.1088/0034-4885/67/8/r03 (2004).
18  Dieterich, J. H. A constitutive law for rate of earthquake production and its application to earthquake clustering. *J. Geophys. Res.* **99**, 2601-2618 (1994).
19  Das, S. & Scholz, C. Theory of time-dependent rupture in the earth. *J. Geophys. Res.* **86**, 6039-6051 (1981).
20  Wilcock, W. S. D., Archer, S. D. & Purdy, G. M. Microearthquakes on the Endeavour segment of the Juan de Fuca Ridge. *Journal of Geophysical Research-Solid Earth* **107**, doi:10.1029/2001jb000505 (2002).
21  Sohn, R. A., Hildebrand, J. A. & Webb, S. C. A microearthquake survey of the high-temperature vent fields on the volcanically active East Pacific Rise (9 degrees 50'N). *Journal of Geophysical Research-Solid Earth* **104**, 25367-25377, doi:10.1029/1999jb900263 (1999).
22  Waldhauser, F. & Tolstoy, M. Seismogenic structure and processes associated with magma inflation and hydrothermal circulation beneath the East Pacific Rise at 9 degrees 50 ' N. *Geochemistry Geophysics Geosystems* **12**, doi:10.1029/2011gc003568 (2011).
23  Tolstoy, M., Waldhauser, F., Bohnenstiehl, D. R., Weekly, R. T. & Kim, W. Y. Seismic identification of along-axis hydrothermal flow on the East Pacific Rise. *Nature* **451**, 181-U187, doi:10.1038/nature06424 (2008).
24  Stroup, D. F. et al. Systematic along-axis tidal triggering of microearthquakes observed at 9 degrees 50 ' N East Pacific Rise. Geophysical Research Letters 36, doi:10.1029/2009gl039493 (2009).
25  Einarsson, P. & Brandsdottir, B. Earthquakes in the Myrdalsjokull area, Iceland, 1978-1985: Seasonal correlation and connection with volcanoes. *Jokull* **49**, 59-73 (2000).
26  Albino, F., Pinel, V. & Sigmundsson, F. Influence of surface load variations on eruption likelihood: application to two Icelandic subglacial volcanoes, Grimsvotn and Katla. *Geophysical Journal International* **181**, 1510-1524, doi:10.1111/j.1365-246X.2010.04603.x (2010).
27  Toda, S., Stein, R. S., Reasenberg, P. A., Dieterich, J. H. & Yoshida, A. Stress transferred by the 1995 M-w = 6.9 Kobe, Japan, shock: Effect on aftershocks and future earthquake probabilities. *Journal of Geophysical Research-Solid Earth* **103**, 24543-24565, doi:10.1029/98jb00765 (1998).
28  Belardinelli, M. E., Cocco, M., Coutant, O. & Cotton, F. Redistribution of dynamic stress during coseismic ruptures: Evidence for fault interaction and earthquake triggering. *Journal of Geophysical Research-Solid Earth* **104**, 14925-14945, doi:10.1029/1999jb900094 (1999).
29  Embley, R. W., Murphy, K. M. & Fox, C. G. High-resolution studies of the summit of Axial Volcano. *Journal of Geophysical Research-Solid Earth and Planets* **95**, 12785-12812, doi:10.1029/JB095iB08p12785 (1990).
30  Chase, R. *et al.* Hydrothermal vents on an axis seamount of the Juan de Fuca ridge. *Nature* **313**, 212-214 (1985).
31  Beeler, N. M. & Lockner, D. A. Why earthquakes correlate weakly with the solid Earth tides: Effects of periodic stress on the rate and probability of earthquake occurrence. *Journal of Geophysical Research-Solid Earth* **108**, doi:10.1029/2001jb001518 (2003).

32. Carpenter, B. M., Collettini, C., Viti, C. & Cavallo, A. The influence of normal stress and sliding velocity on the frictional behaviour of calcite at room temperature: insights from laboratory experiments and microstructural observations. *Geophysical Journal International* **205**, 548-561, doi:10.1093/gji/ggw038 (2016).
33. Chen, J., Verberne, B. A. & Spiers, C. J. Effects of healing on the seismogenic potential of carbonate fault rocks: Experiments on samples from the Longmenshan Fault, Sichuan, China. *Journal of Geophysical Research-Solid Earth* **120**, 5479-5506, doi:10.1002/2015jb012051 (2015).
34. Ikari, M. J., Ito, Y., Ujiie, K. & Kopf, A. J. Spectrum of slip behaviour in Tohoku fault zone samples at plate tectonic slip rates. *Nature Geoscience* **8**, 870-+, doi:10.1038/ngeo2547 (2015).
35. Atkinson, B. K. Subcritical crack growth in geological materials. *J. Geophys. Res.* **89**, 4077-4114 (1984).
36. Shearer, P. M., Prieto, G. A. & Hauksson, E. Comprehensive analysis of earthquake source spectra in southern California. *Journal of Geophysical Research-Solid Earth* **111**, doi:10.1029/2005jb003979 (2006).
37. Boyd, O. S., McNamara, D. E., Hartzell, S. & Choy, G. Influence of Lithostatic Stress on Earthquake Stress Drops in North America. *Bulletin of the Seismological Society of America* **107**, 856-868, doi:10.1785/0120160219 (2017).
38. Dieterich, J. in *Earthquake Source Mechanics. AGU Geophys. Mono.* (eds S. Das, J. Boatwright, & C. Scholz) 37-49 (American Geophysical Union).
39. Roy, M. & Marone, C. Earthquake nucleation on model faults with rate- and state-dependent friction: Effects of inertia. *Journal of Geophysical Research-Solid Earth* **101**, 13919-13932, doi:10.1029/96jb00529 (1996).
40. Stein, R. S. The role of stress transfer in earthquake occurrence. *Nature* **402**, 605-609 (1999).
41. van der Elst, N. J. & Brodsky, E. E. Connecting near-field and far-field earthquake triggering to dynamic strain. *Journal of Geophysical Research-Solid Earth* **115**, doi:10.1029/2009jb006681 (2010).
42. Hill, D. P. e. a. Seismicity Remotely Triggered By the Magnitude 7.3 Landers, California, Earthquake. *Science* **260**, 1617-1623 (1993).
43. Husen, S., Taylor, R., Smith, R. B. & Healser, H. Changes in geyser eruption behavior and remotely triggered seismicity in Yellowstone National Park produced by the 2002 M 7.9 Denali fault earthquake, Alaska. *Geology* **32**, 537-540, doi:10.1130/g20381.1 (2004).
44. Prejean, S. G. *et al.* Remotely triggered seismicity on the United States west coast following the M-W 7.9 Denali fault earthquake. *Bulletin of the Seismological Society of America* **94**, S348-S359, doi:10.1785/0120040610 (2004).
45. Brodsky, E. E., Roeloffs, E., Woodcock, D., Gall, I. & Manga, M. A mechanism for sustained groundwater pressure changes induced by distant earthquakes. *Journal of Geophysical Research-Solid Earth* **108**, doi:10.1029/2002jb002321 (2003).
46. Linde, A. T., Sacks, I. S., Johnston, M. J. S., Hill, D. P. & Bilham, R. G. Increased Pressure From Rising Bubbles As a Mechanism For Remotely Triggered Seismicity. *Nature* **371**, 408-410 (1994).

47	Manga, M. *et al.* Changes in permeability caused by dynamic stresses: Field observations, experiments, and mechanisms. *Reviews of Geophysics* **50**, doi:10.1029/2011rg000382 (2012).
48	Vera, E. et al. The structure of 0-to 0.2-my-old oceanic crust at 9° N on the East Pacific Rise from expanded spread profiles. Journal of Geophysical Research: Solid Earth 95, 15529-15556 (1990).

**Figure Captions**

Figure 1. Cross-section of seismicity preceding the 2015 eruption at Axial Volcano. Red curve is the roof of the axial magma chamber.

Figure 2. Histogram of earthquakes plotted vs. the phase of the vertical component of the tidal stress, in which 0° is the maximum low tide.

Figure 3. Distribution of Coulomb stress changes on 67° dipping normal faults near the axial magma chamber. Positive values favors fault slip, negative inhibit it. For, A) an overpressure of 1 MPa within the magma chamber, and B) a decrease in vertical stress equivalent to a reduction in water level of 1 m. In A) the friction coefficient on the faults $\mu=0.8$, in B) it is the effective friction $\mu'=0.4$. The bulk modulus of the rock is assumed to be $K_r=55$ GPa, and in B) the bulk modulus of the magma $K_m=1$ GPa.

Figure 4. Systematics of the magma chamber deformation system. The vertical axis $\chi$ is the average change in $\Delta$CFS on a 67° dipping normal fault from the tip of the magma chamber to the surface, normalized by the vertical tidal stress. The red area defines the conditions in which low tides encourage seismicity and high tides discourage it, and the blue area vice versa.

Figure 5. Normalized seismicity rate change vs. change in Coulomb stress. Blue curve is the rate and state friction version and the red curve is the stress corrosion version.

**Methods**

**Coulomb stress modeling.**

Coulomb stress calculation is performed with the commercial Finite Element Modelling software COMSOL Multiphysics ® (https://www.comsol.com). We use a 100 x 100 x 50 km domain designed to limit boundary effects. Boundaries conditions are zero-displacement for the bottom and lateral boundaries and free-displacement for the top boundary corresponding to the Earth's surface. For the host rock, we assume an isotropic and homogeneous elastic medium with a bulk modulus $K_r$ of 55 GPa and a

Poisson's ratio $\nu_0$ of 0.25, which is in accordance with seismic velocities recorded on the East Pacific Rise[48]. At Axial Seamount, multichannel seismic-reflection has inferred a 14-km long by 3-km-wide shallow magma reservoir located at 1.1-2.3 km depth[6,11]. We therefore model the magma reservoir as a 3D ellipsoid cavity with semi-axis: a=7 km, b=1.5 km, and c=0.5 km, and top depth located at 2km below the surface. In our modeling, the initial stress field is lithostatic and stress perturbations are calculated considering two scenarios: (1) the pressurization of the magma reservoir and (2) the effect of ocean tides. For the first scenario, the overpressure inside the reservoir is modeled by applying a constant normal stress applied the boundary of the ellipsoid. For the second scenario, the stress changes due to ocean low tides are modeled by applying a boundary load at the surface corresponding to a 1 m decrease in the water level. Surface unloading causes the reservoir expansion resulting in a magma pressure change, which depends on the reservoir volume, the bulk modulus of the magma and the elastic properties of the host rock. The pressure change is applied on the reservoir's wall considering different bulk modulus $K_m$ from 0 to 12 GPa. For each model, the Coulomb failure stress change is calculated on specific fault planes using $\Delta CFS = \Delta\tau - \mu\Delta\sigma$, where $\Delta\sigma$ is the normal stress change, $\Delta\tau$ the tangential stress changes and $\mu$ the friction coefficient.

**Seismicity catalog**

The earthquake catalog is the same as in Wilcock (2016) and is archived in the Interdisciplinary Earth Data Alliance Marine Geoscience Data System (DOI:10.1594/IEDA/322843).

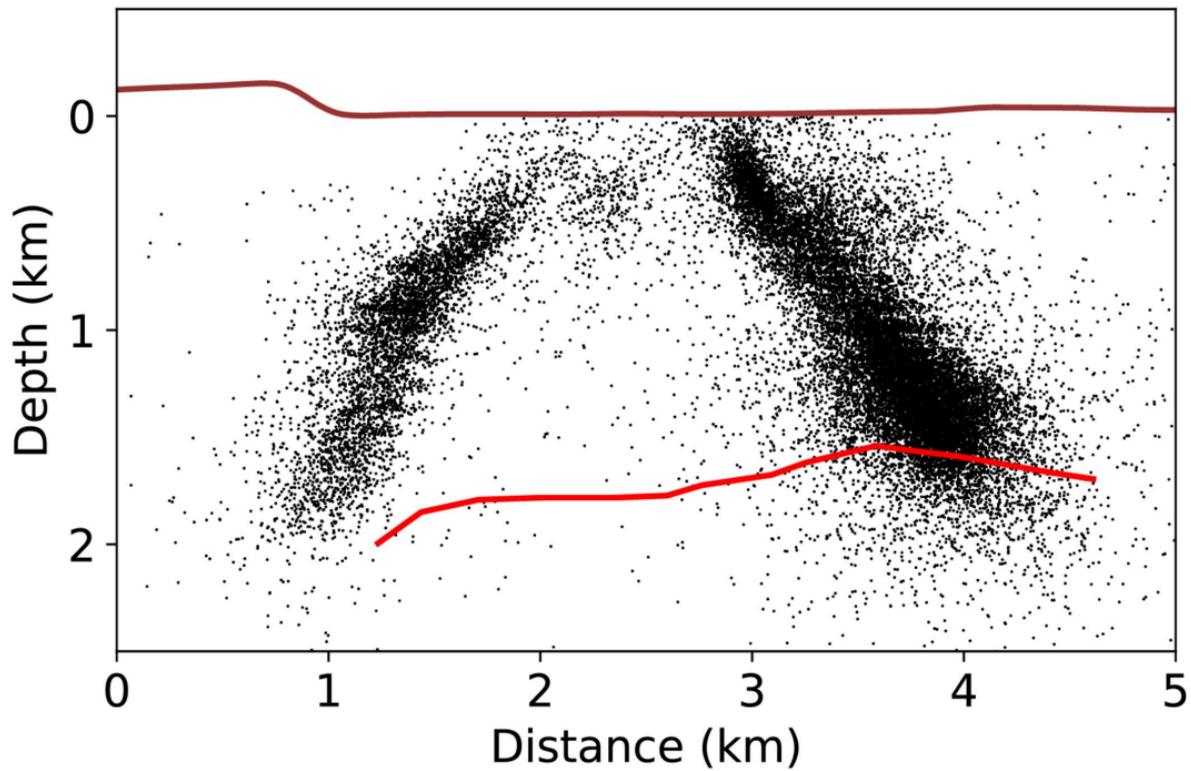

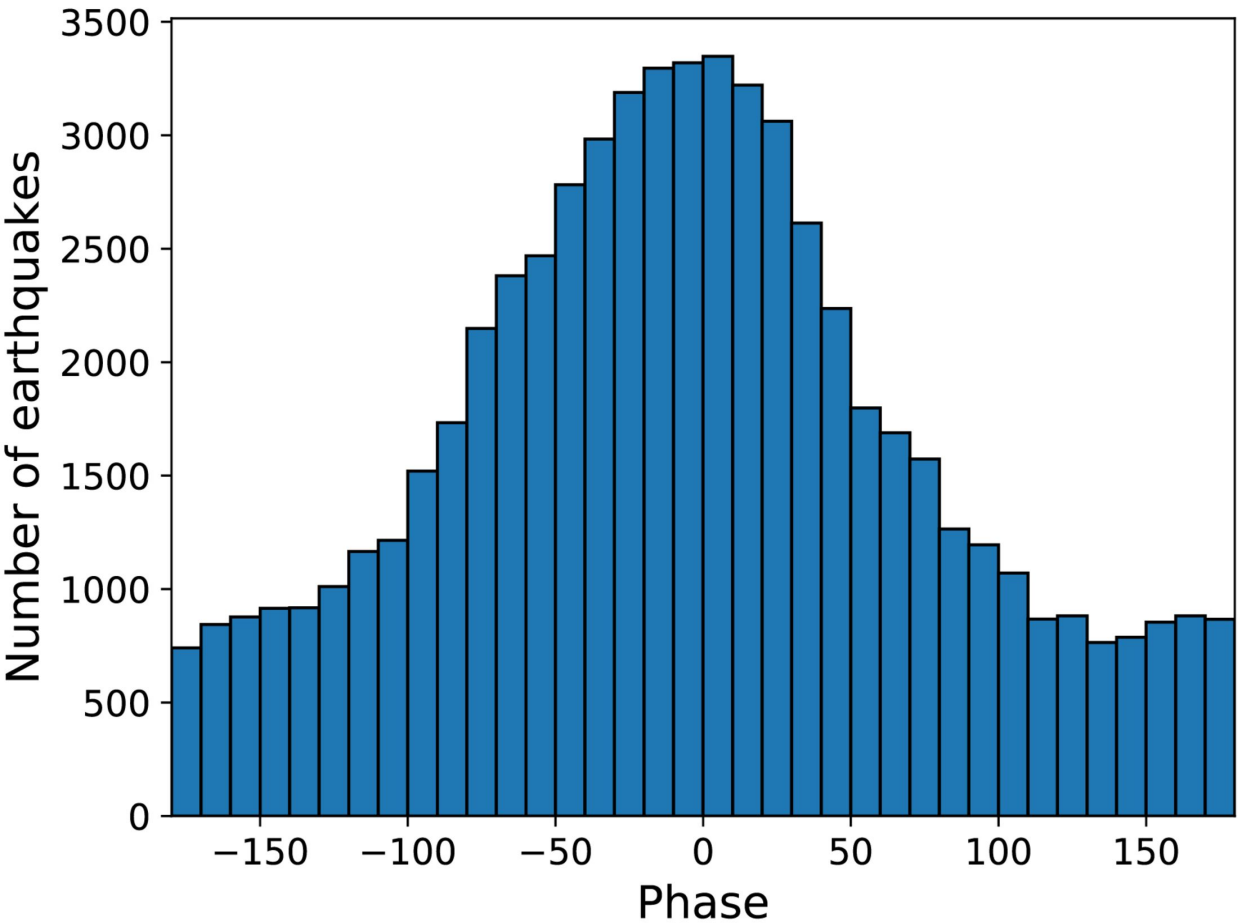

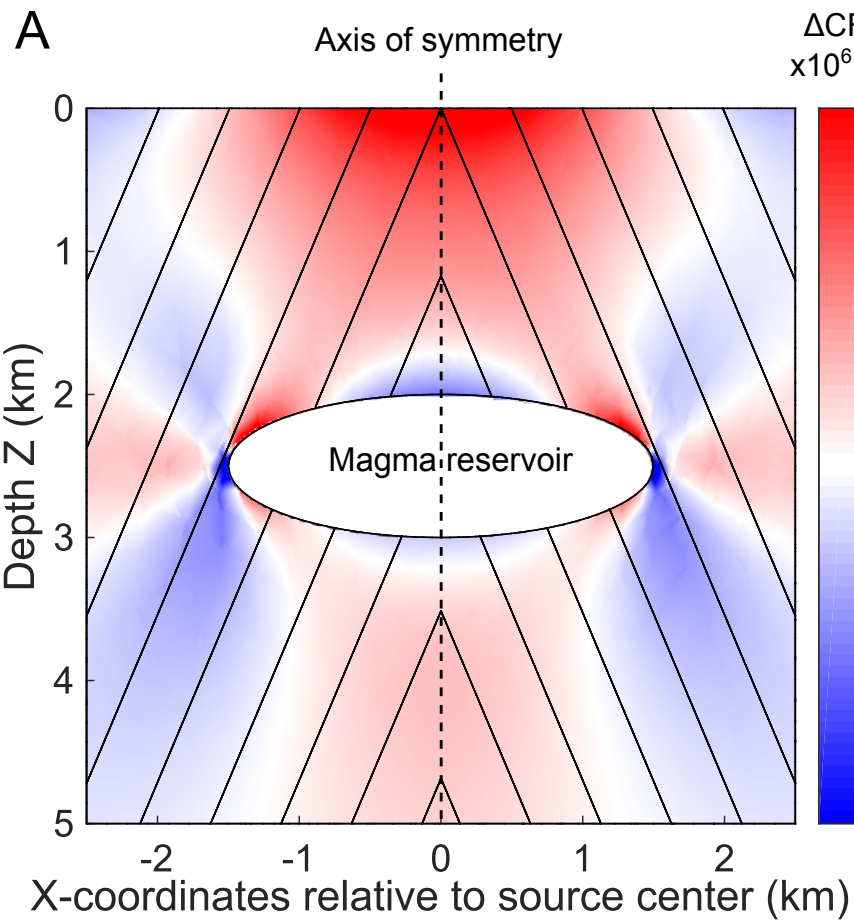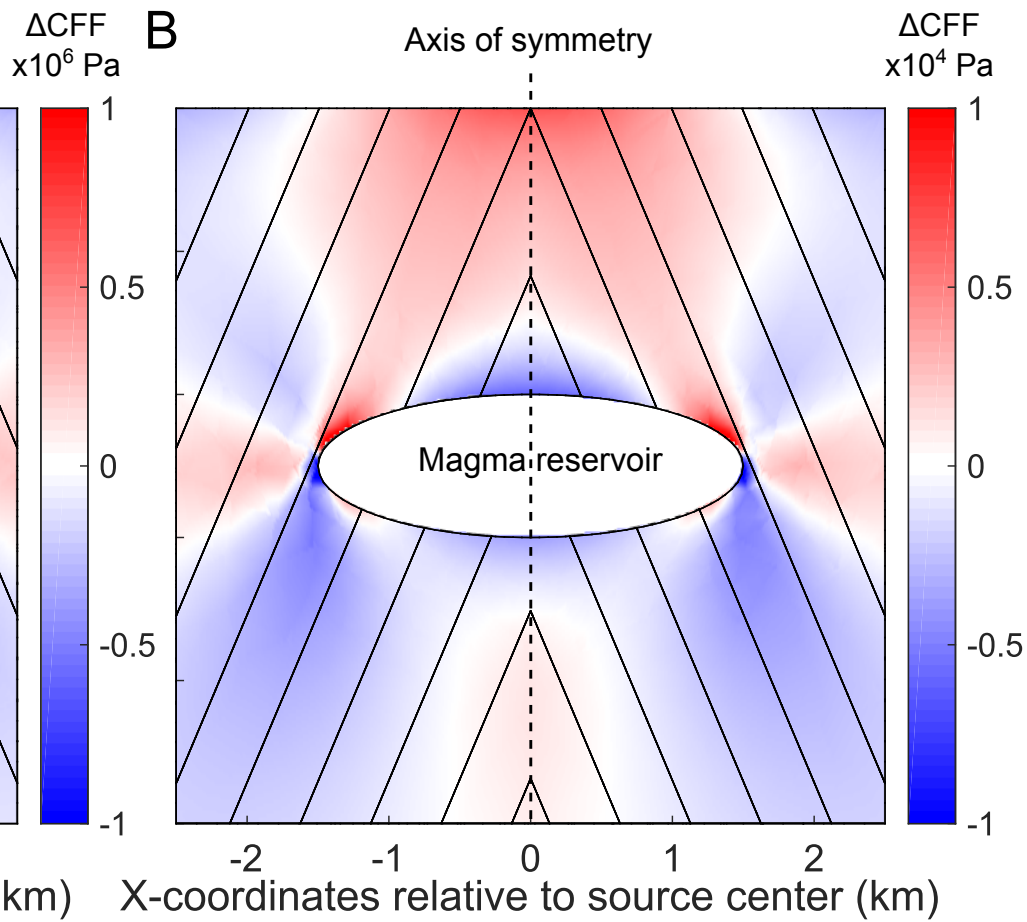

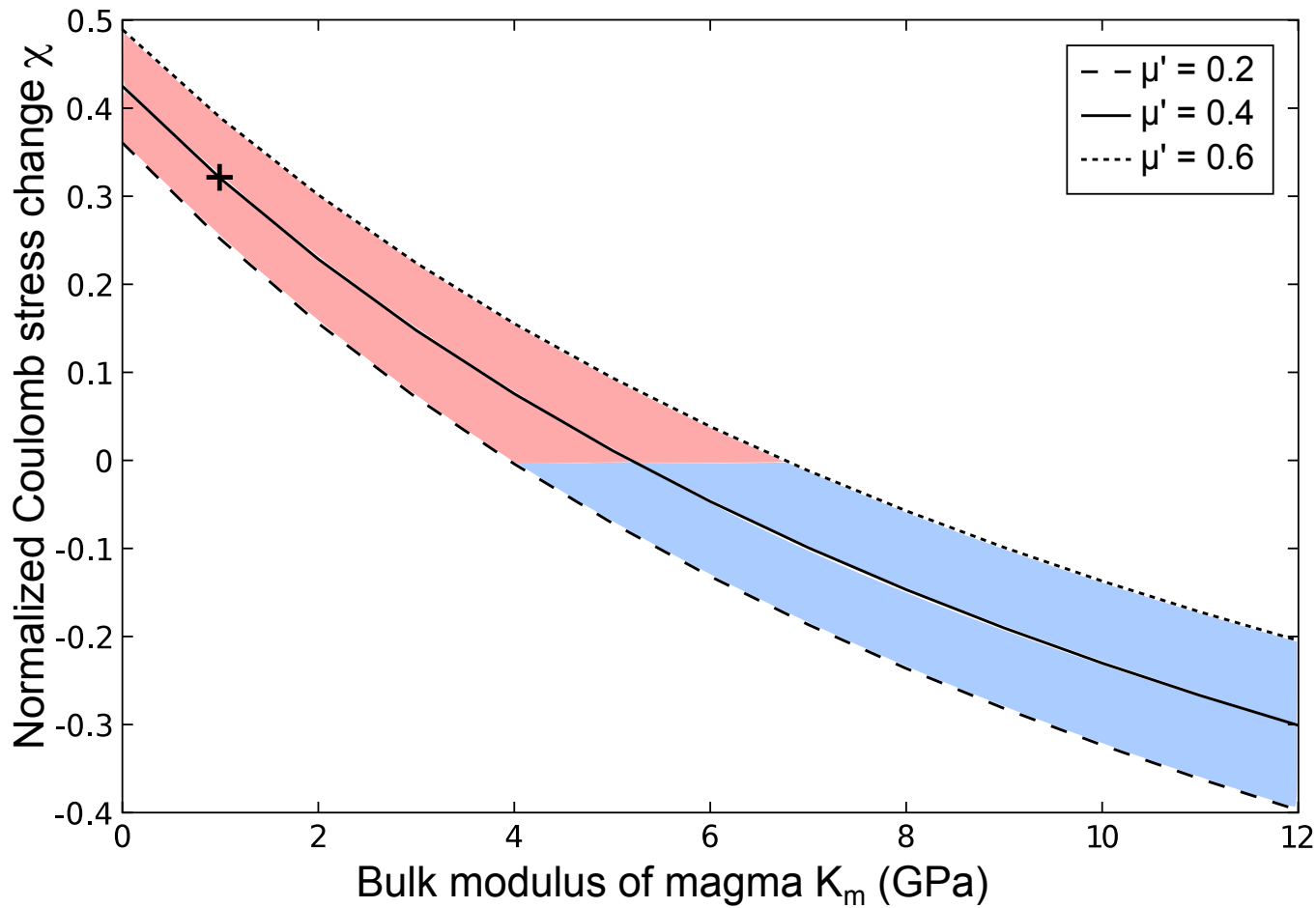

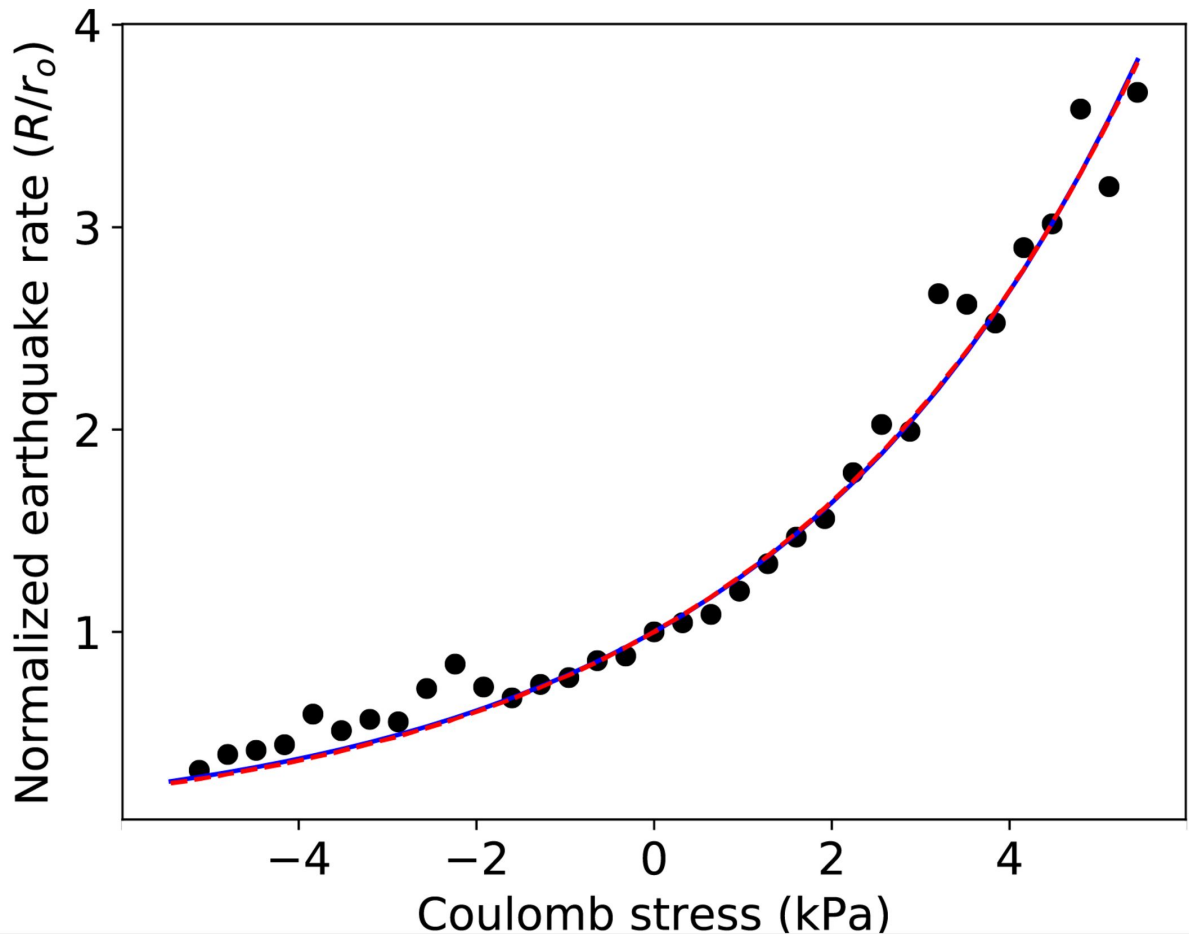